\documentclass[a4paper]{jpconf}
\usepackage[utf8x]{inputenc}
\usepackage{graphicx}
\usepackage{multirow}
\usepackage{url}

\begin{document}
\title{Five-loop numerical evaluation of critical exponents of the $\varphi^4$ theory}
\author{L.Ts. Adzhemyan,  M.V. Kompaniets}
\address{Department of Theoretical Physics, St.~Petersburg State University,
Uljanovskaja 1, St.~Petersburg, Petrodvorez, 198504 Russia}
\ead{mkompan@gmail.com}

\begin{abstract}
We present a new approach to calculation of anomalous dimensions in the framework
 of $\epsilon$-expansion and renormalization group method. This approach allows one to skip the calculation 
of renormalization constants and express anomalous dimensions in terms of renormalized 
diagrams, which are presented in a form suitable for numerical calculations.
This approach can be easily automated and extended to a wide range of models.
The power of this approach is illustrated on 5 loop calculations of beta-function 
and anomalous dimensions in $\phi^4$ model. 
\end{abstract}

The renormalization group approach is one of the most effective methods of quantum field theory.
The most studied in the framework of this approach is the model $\varphi^4$. On the one hand, this model describes the second order phase transition for different systems \cite{vasiliev,Zinn,phi47}, on the other hand, it is usually used as a testing ground for approbation of new methods of Feynman diagrams calculations. In particular, in  the framework of this model were tested the methods like Gegenbauer polynomial x-space technique (GPTX) \cite{gegenbauers,phi412},
integration-by-parts method \cite{ibp} and infrared $R^*$-operation \cite{Rstar}.
These methods allowed to reach maximal accuracy in critical exponents (5th order of $\varepsilon$-expansion) available now \cite{phi412,phi434,phi456}. Critical exponent $\eta$ was calculated with 5-loop accuracy in \cite{phi412}, then  calculation of $\beta$-function in 5th order of perturbation theory  was performed in \cite{phi434}. Later on some inaccuracies  were found in this calculation \cite{phi456} and results for index $\eta$ and $\beta$-function were corrected. In calculations \cite{phi456} the same method, namely $R^*$-operation technique, as in \cite{phi434} was used, that is why \cite{phi456} cannot be treated as fully independent check of the preceding works \cite{phi412,phi434}.

Nowadays methods \cite{gegenbauers, ibp, Rstar} are successfully applied to different models of quantum field theory, however application of these methods to the models of critical dynamics \cite{vasiliev,folk} (which are the main object of our interest) encounters significant problems due to some specific features of these models. In \cite{tmf1,tmf2} the new approach to calculating of critical exponents was proposed.
This approach allows one to skip the calculation of renormalization constants and express anomalous dimensions in terms of renormalized diagrams, which are presented in a form suitable for numerical calculations. The main benefit of this approach is that it can be easily automated and extended to a wide range of models.

In order to test our approach we performed 5-loop calculation of critical exponents. On the one hand it allows us to examine the efficiency of our approach in high orders of perturbation theory, and on the other hand approach discussed in this paper deals nothing with the $R^*$-operation and can be used as a fully independent check of the previous results \cite{phi412, phi434,phi456}.

The results obtained for model $\varphi^4$ using our approach ($d=4-\varepsilon$)
$$ 
\eta=  0.0185185185\;\epsilon^2 + 0.0186900(6)\epsilon^3   - 0.0083286(2)\epsilon^4  + 0.025656(2)\epsilon^5 
$$
$$
\omega = \epsilon - 0.62962963(8)\epsilon^2 + 1.6182211(7)\epsilon^3  - 5.23513(2)\epsilon^4 + 20.7499(9)\epsilon^5
$$
are in good agreement with the results obtained in \cite{phi456} 
$$
\eta= 0.0185185185\;\epsilon^2 + 0.0186899862\;\epsilon^3   - 0.008328770\;\epsilon^4 + 0.025656451\;\epsilon^5
$$
$$
\omega =  \epsilon- 0.629629629\;\epsilon^2  + 1.61822067\;\epsilon^3 - 5.2351359\;\epsilon^4 +20.74984\;\epsilon^5 
$$
and confirm inaccuracies found in \cite{phi412,phi434}.

The plan of the paper is as follows. In the next section we shortly describe our approach and renormalization scheme used.
In the second section we present examples of diagram calculation in discussed approach. The last section is devoted to comparison of our results with results of \cite{phi412,phi434, phi456} in more details.

\section{Normalization point approach}
Let us consider $\varphi^4$ theory in euclidean space with $d=4-\varepsilon$ dimensions.
Basic action for this theory has the following form \cite{vasiliev}
\begin{equation}\label{p0}
 S_B=-\frac{1}{2}m^2 \varphi^2-\frac{1}{2}\left(\partial\varphi \right)^2-\frac{1}{4!} g\mu^\varepsilon \varphi^4.
\end{equation}
Diagrams of Green functions $\Gamma_i$ calculated in the theory (\ref{p0}) have ultraviolet divergences, these divergences can be removed by adding counterterms to action (\ref{p0}).
This leads to renormalized action of the following form
\begin{equation}\label{p1}
 S=-\frac{1}{2}(m^2 Z_1+\delta m^2)\varphi^2-\frac{1}{2}Z_2\left(\partial\varphi \right)^2-\frac{1}{4!}Z_3 g\mu^\varepsilon \varphi^4,
\end{equation}
where renormalization constants $Z_1, Z_2, Z_3$ are expressed in terms of renormalization constants of mass $Z_{m^2}$, field $Z_{\varphi}$ 
and coupling constant $Z_{g}$ by the relations
\begin{equation}\label{ZZ}
Z_1=Z_{m^2}Z_{\varphi}^2,\qquad Z_2=Z_{\varphi}^2, \qquad Z_3=Z_{g} Z_{\varphi}^4.
\end{equation}
Particular choice of renormalization constants is determined by renormalization scheme. In this work we will use normalization point (NP) scheme defined as follows.
The mass shift $\delta m^2$  is determined by the condition 
\begin{equation}\label{usl}
 \Gamma_2^R\mid_{p=0,m=0}=0\,.
\end{equation}
Renormalization constants $Z_i$ are defined in such a way that at normalization point  $p=0, \,\mu=m$ the renormalized one-particle irreducible (1-PI) two-point function $\Gamma_2^R$ and 1-PI four-point function $\Gamma_4^R$ are equal to their loopless terms:
\begin{equation}\label{usl1}
  \Gamma_2^R\mid_{p=0,\mu=m}=-m^2\,, \quad \partial_{p^2}\Gamma_2^R\mid_{p=0,\mu=m}=-1\,, \quad \Gamma_4^R\mid_{p=0,\mu=m}=-g m^{2\varepsilon}\,.
\end{equation}
Henceforth it is convenient to use normalized functions
\begin{equation}\label{123}
\bar{\Gamma}_1=-\left(\frac{\Gamma_2-\Gamma_2|_{m=0}}{m^2} \right) \,, \quad \bar{\Gamma}_2=-\partial_{p^2}\Gamma_2\,, \quad
 \bar{\Gamma}_4=\Gamma_4/(-g\mu^{2\varepsilon})\,,
\end{equation}
which, according to (\ref{usl}), (\ref{usl1}), satisfy the following relations
\begin{equation}\label{usl2}
\bar \Gamma_2^R|_{p=0, m=0} =0,\quad \bar \Gamma_1^R|_{p=0, \mu=m} =1, \quad \bar \Gamma_2^R|_{p=0, \mu=m} =1, \quad \bar \Gamma_4^R|_{p=0, \mu=m} =1 .
\end{equation}
Renormalization constants defined by conditions (\ref{usl2}) do not depend on $m$, like in minimal subtraction (MS) scheme. Renormalization group equations also coincide with that in MS scheme:
\begin{equation}\label{RG}
\left(  {\cal{D}}_\mu+\beta\partial_g-\gamma_{m^2}{\cal{D}}_{ m^2} \right) \Gamma_i^R=n\gamma_\varphi \Gamma_i^R,
\end{equation}
where ${\cal{D}}_{ m^2}\equiv m^2 \partial_{m^2}\mid_{\mu,g}, {\cal{D}}_{ \mu}\equiv \mu \partial_{\mu}\mid_{m,g},\, \gamma_i=\beta \partial_g \ln Z_i,\, \beta=-g(\varepsilon+\gamma_g)$. Using (\ref{RG}), one can obtain for normalized functions (\ref{123})
\begin{equation}\label{RG1}
\left(  {\cal{D}}_\mu+\beta\partial_g-\gamma_{m^2}{\cal{D}}_{ m^2} \right) \bar{\Gamma}_i^R=\gamma_i \bar{\Gamma}_i^R,
\end{equation}
where, according to (\ref{ZZ}),
\begin{equation}\label{RG2}
 \gamma_1=\gamma_{m^2}+2\gamma_\varphi,   \qquad \gamma_2=2\gamma_\varphi,  \qquad \gamma_3=\gamma_g+4\gamma_\varphi.
\end{equation}

Let us consider eqs. (\ref{RG1}) at normalization point $p=0, \mu=m$.
Taking into account (\ref{usl2}) we can express RG-functions $\gamma_i$ in terms of renormalized functions $\bar \Gamma_i^R$ at normalization point \cite{tmf1, tmf2}:
\begin{equation}\label{F}
 \gamma_i=\frac{2F_i}{1+F_2-F_1}\,, \qquad  F_i\equiv \left( - m^2\partial_{m^2}\bar\Gamma_i^R\right)|_{p=0, \mu=m} \,,\qquad i=1,\,2,\,\,4 \,.
\end{equation}

For renormalized Green functions $\Gamma_i^R$ it is possible to replace addition of counterterms (renormalization constants) by R-operation acting on Green functions $\Gamma_i$ calculated in the theory (\ref{p0}):
\begin{equation}\label{2d}
\Gamma_i^R= R\Gamma_i.
\end{equation}
Using R-operation let us define the following functions
\begin{equation}\label{f}
 f_i= R\left[\left (- m^2\partial_{m^2}\bar\Gamma_i\right) \right]|_{p=0, \mu=m} \,.
\end{equation}
It is shown in \cite{tmf1,tmf2} that interrelations between these functions and functions $F_i$ (\ref{F}) are
\begin{equation}\label{Ff}
 f_i-F_i=f_i F_1\,, \qquad i=2,\,4.
\end{equation}
This allows us to rewrite (\ref{F}) in the following form
\begin{equation}\label{00}
 \gamma_i=\frac{2f_i}{1+f_2}\,, \qquad i=2,\,\,4 \,.
\end{equation}
Relations (\ref{f}) and (\ref{00}) will be used in further numerical calculations. The advantage of these relations with respect to (\ref{F}) is that R-operation in (\ref{f}) is taken at normalization point, and thus has more simple form. 
R-operation in (\ref{f}) can be expressed in terms of product of $1-K_i$ operations, which removes all divergences from diagrams \cite{Zavialov}
\begin{equation}\label{2d}
R\Gamma=\prod_i (1-K_i)\Gamma,
\end{equation}
where product is taken over all relevant (UV-divergent) subgraphs of a particular diagram, including diagram as whole.
Differentiation of diagram lines in (\ref{f}) with respect to squared mass ($-\partial_{m^2}$) is equivalent to insertion of unit vertex into a line (``dot insertion''). That is why in such a graphs there are quadratically divergent 1-PI two-point  subgraphs and logarithmically divergent 1-PI four-point  subgraphs and two-point  subgraphs with inset dot.

In renormalization scheme under consideration a subtraction operation for graphs taken at $\mu=m$ is equivalent to the subtraction of initial part of Taylor series over external momenta
 \begin{equation}\label{v}
 (1-K_i)F(k)=F(k)-\sum_{m=0}^n \frac{k^m}{m!} F^{(m)}|_{k=0},
\end{equation} 
where $n=0$ for logarithmically divergent subgraphs and $n=2$ for quadratically divergent ones. 
It is possible to rewrite equation (\ref{v}) using representation for reminder of Taylor series
%
\begin{equation}\label{VV}
 (1-K_i)F(k)=\frac{1}{n!}\int_0^1 da (1-a)^n \partial_a^{n+1}F(ak).
\end{equation} 
Combining (\ref{2d}) and (\ref{VV}) we can construct the following representation for $R$-operation \cite{Zavialov}
\begin{equation}\label{Vd}
R\chi=\prod_i \frac{1}{n_i!}\int_0^1 da_i (1-a_i)^{n_i} \partial_{a_i}^{n_i+1}\chi(\{a\}),
\end{equation}
where product is taken over all 1-PI  divergent subgraphs $\chi_{(i)}$ (including diagram $\chi$ as whole) with canonical dimension $ n_i\geq 0$, and $a_i$ -- parameter that stretches moments flowing into $i$-th subgraph. The main advantage of such a representation is that the renormalized function is expressed by integrals finite at $\varepsilon=0$, and there is no cancellation of large terms in integrand (``theory without divergences'' \cite{Zavialov}).

\section{Example of a diagram calculation}
Let us consider 2-loop contribution to $f_4$ from (\ref{f}) determined by 2-loop four-point diagram of  $\Gamma_4$ 
\begin{equation}
  \includegraphics[width=3cm]{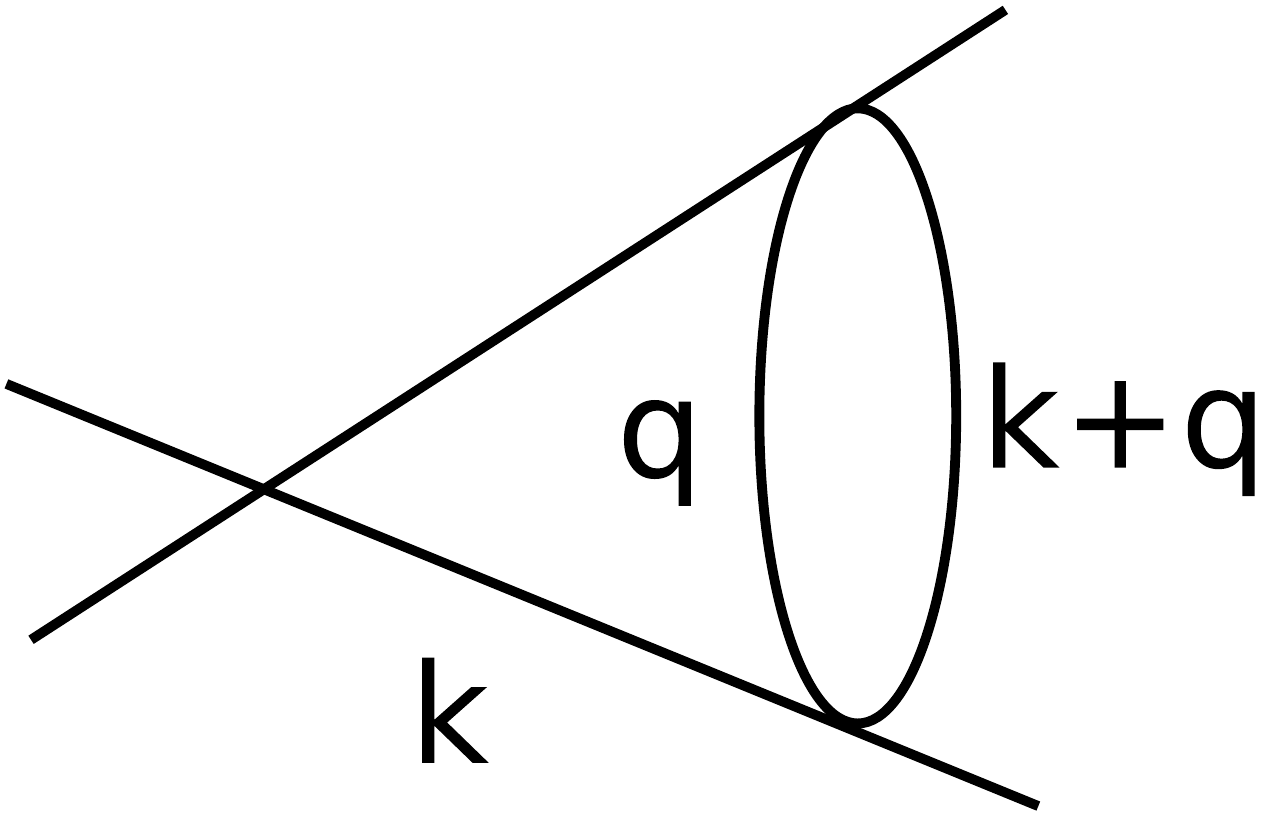}\label{D1}.
\end{equation}
The integral that corresponds to the diagram (\ref{D1}) is
\begin{equation}\label{J}
J=\int\frac{d{\bf k}}{(2\pi)^d}\int\frac{d{\bf q}}{(2\pi)^d} \,I({\bf k},{\bf q})\,,
\end{equation}
where
\begin{equation}\label{I}
 I({\bf k},{\bf q})=\frac{1}{(k^2+m^2)^2}\cdot\frac{1}{q^2+m^2}\cdot\frac{1}{({\bf k}+{\bf q})^2+m^2}\,.
\end{equation}
At $\varepsilon \rightarrow 0$ there is logarithmic superficial divergence in the diagram and logarithmic divergence in subgraph. According to (\ref{f}) contribution of this diagram to function $f_4$ is given by
\begin{equation}
\delta f_4 = 3\, g^2 m^{2\varepsilon} R (-m^2\partial_{m^2} J), \label{df}
\end{equation}
here multiplier 3 is a symmetry coefficient for diagram (\ref{D1}). Differentiation with respect to $m^2$ removes superficial divergence. Performing differentiation over $m^2$ and taking into account diagram symmetry we obtain two terms for this contribution
$$
\delta f_4 = \delta f_4^{(1)} + \delta f_4^{(2)}
$$
by the diagrams
\begin{equation}
 \includegraphics[width=3cm]{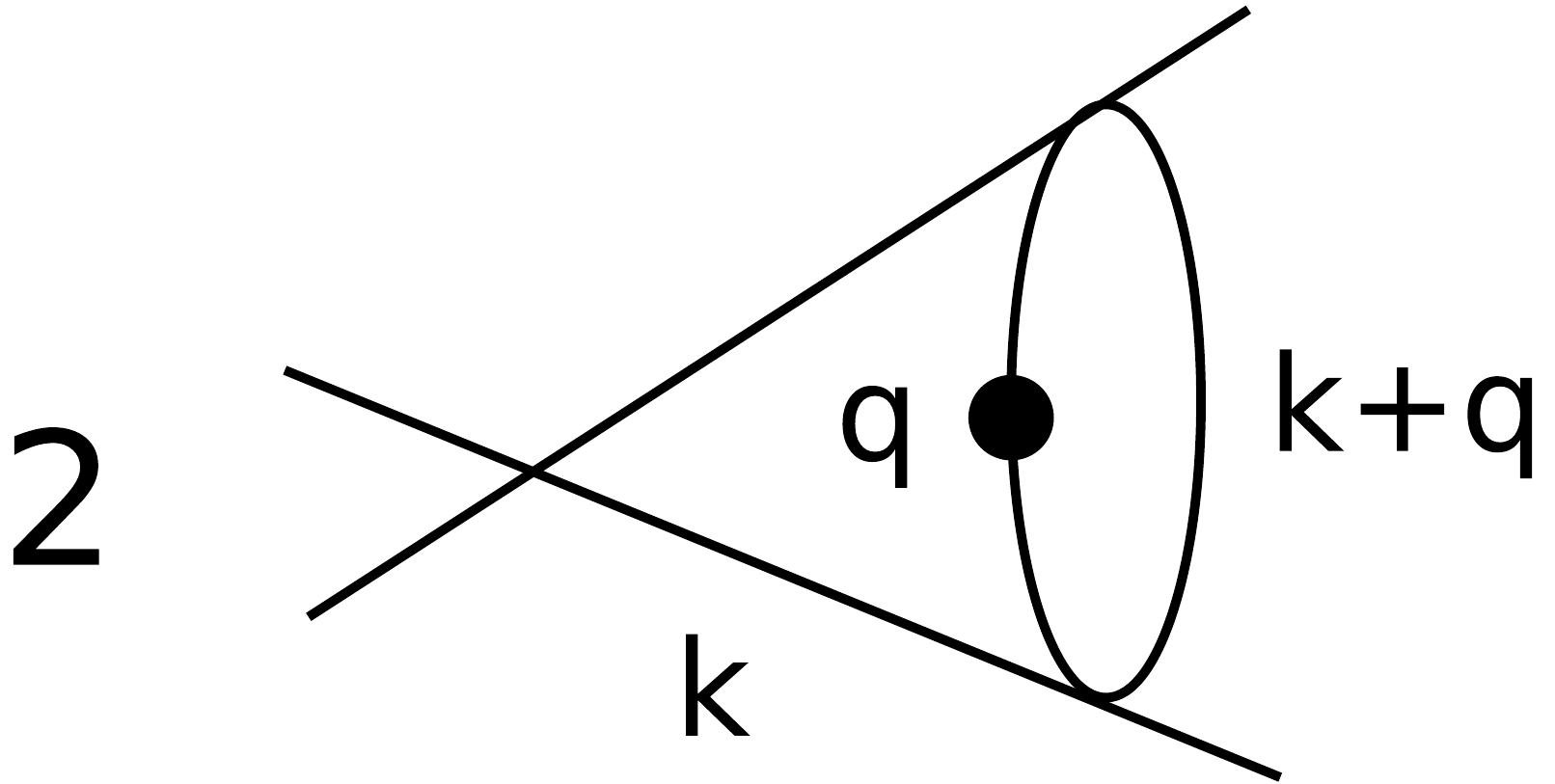}\label{D3},
\end{equation}
\begin{equation}
 \includegraphics[width=3cm]{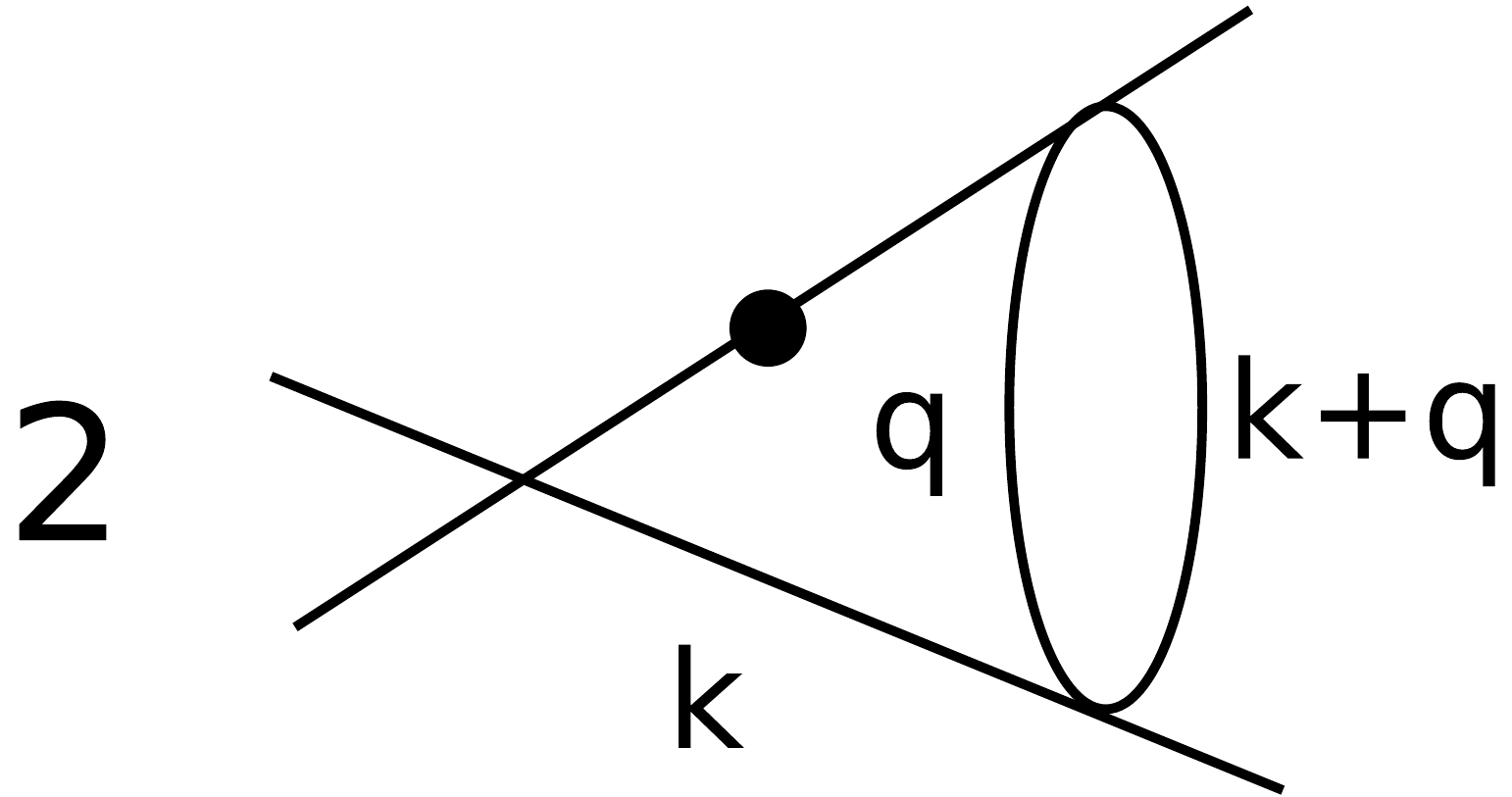}\label{D2},
\end{equation}
with the following integrands in (\ref{J})
\begin{equation}\label{I1}
 I^{(1)}({\bf k},{\bf q})=\,2\frac{1}{(k^2+m^2)^2}\cdot\frac{1}{(q^2+m^2)^2}\cdot\frac{1}{({\bf k}+{\bf q})^2+m^2},
\end{equation}
\begin{equation}\label{I2}
 I^{(2)}({\bf k},{\bf q})=\,2\frac{1}{(k^2+m^2)^3}\cdot\frac{1}{q^2+m^2}\cdot\frac{1}{({\bf k}+{\bf q})^2+m^2}.
\end{equation}
Dot insertion in the diagram (\ref{D3}) removes subgraph divergence. This results in the following: integral (\ref{I1}) is UV-finite and R-operation in (\ref{df}) for this term is trivial ($R\equiv 1$). This allows us to calculate $\delta f_4^{(1)}$ numericaly as Taylor series over $\varepsilon$.

In diagram (\ref{D2}) divergence in subgraph is still present and R-operation is nontrivial. To construct R-operation of type (\ref{Vd}) that will remove divergence in subgraph we must introduce  streching parameter $a$ for momenta flowing into this subgraph
\begin{equation}
  \includegraphics[width=34mm]{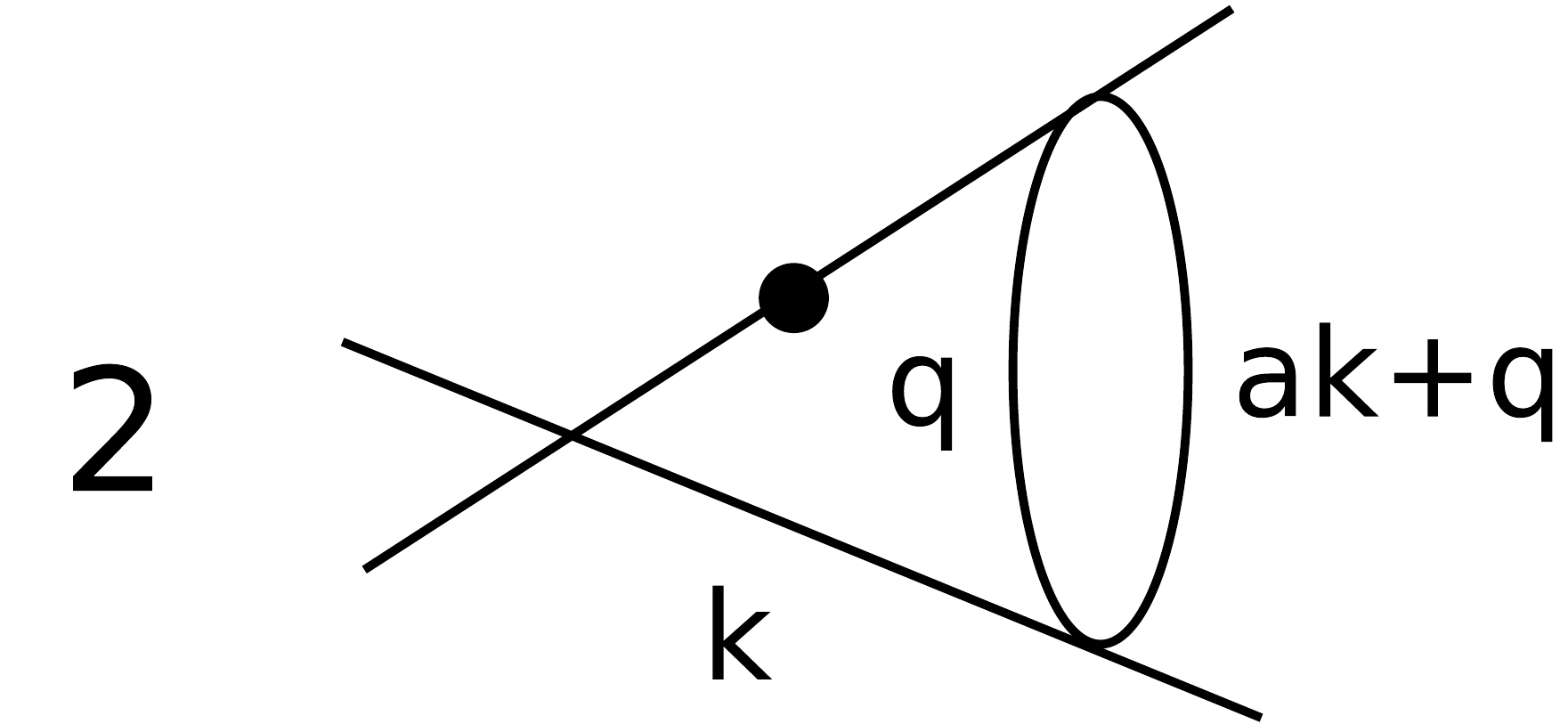}\label{D2a}.
\end{equation}
Applying R-operation (\ref{Vd}) to this diagram we obtain its  contribution  to function $f_4$ (\ref{f}):
 \begin{equation}\label{df4}
 \delta f_4^{(2)}=3\, g^2 m^{2\varepsilon}\int\frac{d{\bf k}}{(2\pi)^d}\int\frac{d{\bf q}}{(2\pi)^d} \int_0^1 da \,\partial_a \,I_2({\bf k},{\bf q},a)\,,
\end{equation}
where
\begin{equation}\label{I2a}
 I^{(2)}({\bf k},{\bf q},a)=\,2\frac{1}{(k^2+m^2)^3}\cdot\frac{1}{q^2+m^2}\cdot\frac{1}{(a{\bf k}+{\bf q})^2+m^2}.
\end{equation}
Performing differentiation with respect to $a$ and transition to dimensionless variables ${\bf k}\rightarrow m{\bf k},\,{\bf q}\rightarrow m{\bf q}$ we obtain
\begin{equation}\label{df4a}
\delta f_4^{(2)}=-3\, g^2 \int\frac{d{\bf k}}{(2\pi)^d}\int\frac{d{\bf q}}{(2\pi)^d} \int_0^1 da\, \frac{1}{(k^2+1)^3}\cdot\frac{1}{q^2+1}\cdot\frac{4ak^2+4{\bf kq}}{\left[ (a{\bf k}+{\bf q})^2+1\right]^2 }.
\end{equation}
This integral is finite at $\varepsilon=0$ and can be evaluated numerically.

In practice for numerical calculations it is more convenient to use Feynman parameters $u_i$. It is well known that it is possible to write integrand in Feynman representation directly from diagram (without necessity of writing momentum representation), it turns out that it may be extended to integrands of type (\ref{I2a}) with streching parameters.

Let us illustrate this on diagram (\ref{D2a})
\begin{equation}
  \includegraphics[width=3cm]{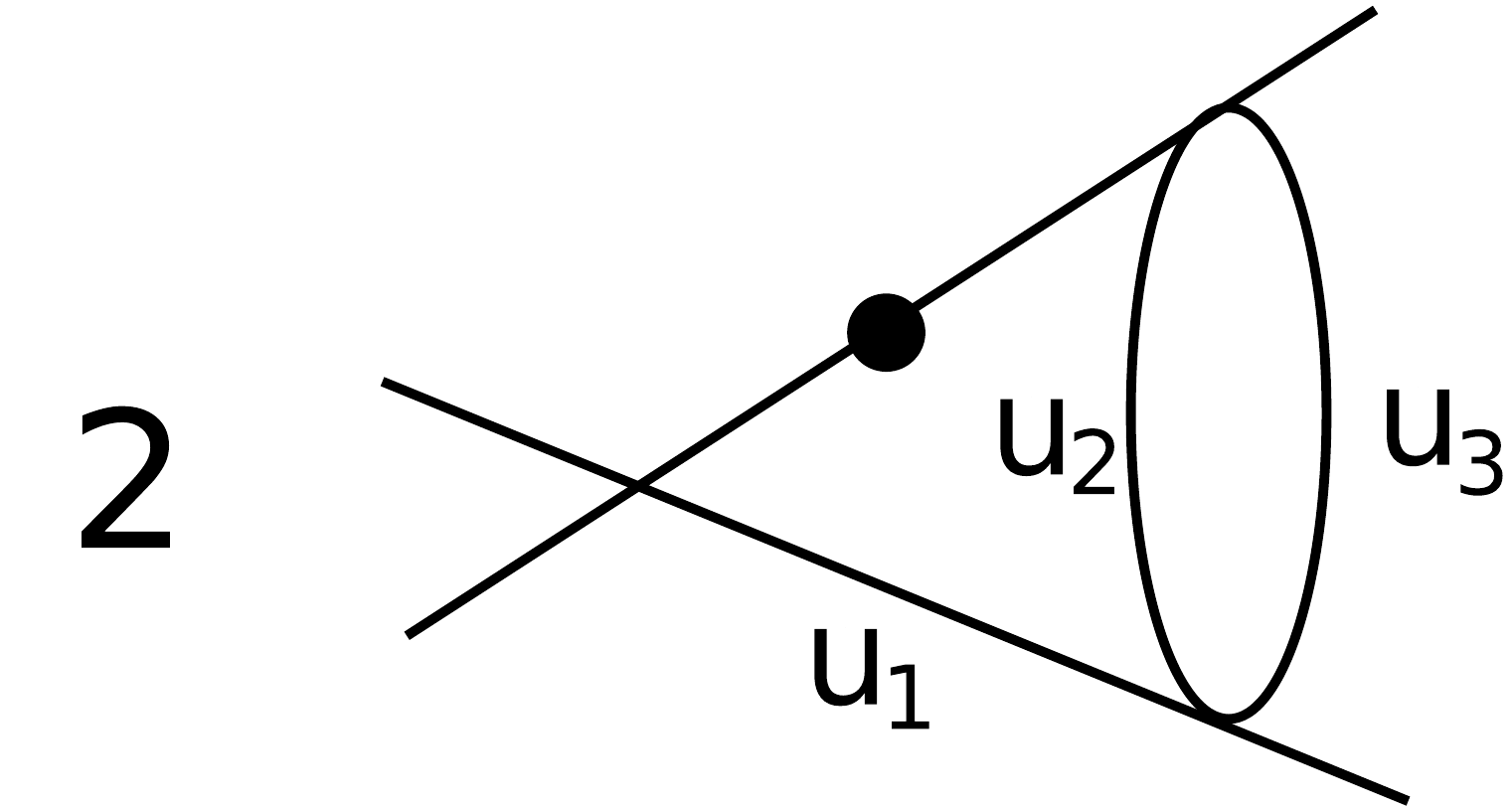}\label{D2b}.
\end{equation}
Contribution $ \delta f_4^{(2)}$ in Feynman representation will have the following form: 
\begin{eqnarray}\label{df4b} \nonumber
 \delta f_4^{(2)}=3\, g^2 \frac{\Gamma(1+\varepsilon)\Gamma^2(2-\varepsilon/2)}{4} \cdot\, J_4\,, \\ 
J_4=2\int_0^1 du_1 \int_0^1 du_2 \int_0^1 du_3 \int_0^1 da \, \partial_a \frac{u_1^2\, \delta(u_1+u_2+u_3-1)}{\left(u_1u_2+u_1u_3+au_2u_3 \right)^{2-\varepsilon/2} }\,.
\end{eqnarray}
This integral is improper, because of denominators zeroes. In the absence of differentiation with respect to streching parameter denominators zero at $u_1=1, u_2=u_3=0$ will produce a pole on $\varepsilon$. But after differentiation over $a$ we got the following expression:
\begin{equation}\label{J4}
 J_4=(\varepsilon-4)\int_0^1 du_1 \int_0^1 du_2 \int_0^1 du_3 \int_0^1 da \frac{u_2u_3u_1^2\, \delta(u_1+u_2+u_3-1)}{\left(u_1u_2+u_1u_3+au_2u_3 \right)^{3-\varepsilon/2} }\,.
\end{equation}
One can see that this singularity becomes integrable and it is possible to calculate this integral as Taylor series on $\varepsilon$.

Presence of integrable singularities does not allow to reach desired accuracy in the numerical calculations. This problem is solved by applying Sector Decomposition technique \cite{heinrich}. In presence of streching parameters it is not possible to use existing sector decomposition strategies, but we've developed generalization of strategy S \cite{strategyS} that can be applied to this type of integrals. This allows us to calculate anomalous dimensions at 5th order of perturbation theory.

\section{Diagram-by-diagram comparison with calculation in MS scheme. 5-loop results. }

For more precise check of the results obtained in \cite{phi456} we performed diagram-by-diagram comparison of our results and results of  \cite{phi456}\footnote{values for particular diagrams are presented in \cite{phi47}}.
To perform diagram-by-diagram comparison  from values calculated in NP scheme we need to construct counterterm contributions in MS scheme. Because of the fact that in NP scheme we calculate only finite parts of diagrams (\ref{f}), first of all we need to construct counterterms  in NP scheme, and then recalculate them from NP scheme to MS one.

The first problem (construction of counterterms in NP scheme) can be solved using representation for couterterms obtained in \cite{tmf1}
\begin{equation}
 Z_i^{(n)} = \frac{2}{n\epsilon}\left({\cal N}\bar\Gamma_i^{(n)}-{\cal J}\bar\Gamma_i^{(n)}\right)\,,
\end{equation}
where $n$ -- number of loops, ${\cal N}\bar\Gamma_i = f_i$, and the second term is defined by sum of diagrams from lower orders of perturbation theory (this reflects the fact that all high order poles can be expressed in terms of lower order diagrams).

The next step is recalculation of counterterms from NP scheme to MS one. It can be performed using so called $R^{-1}$ operation \cite{rminus}. This operation allows one to recursively recalculate counterterms from one renormalization scheme to another (without necessity of calculation of diagrams).

We've performed recalculation of our results to MS scheme and found that results are in good agreement with \cite{phi47,phi456}: all diagrams coincides with at least $10^{-5}$ accuracy. This confirms corrections made in \cite{phi456}.

In \cite{phi456} it is stated that there are inaccuracies in 6 diagrams in \cite{phi412,phi434}. In tables \ref{tab:G2}, \ref{tab:G4} one can find diagrams corrected in \cite{phi456}, values from previous papers \cite{phi412,phi434}, and results of our numerical calculations with error estimations.
\begin{table}[h!]
\caption{\label{tab:G2} Diagrams from two-point Green function calculated in \cite{phi412} and corrected in \cite{phi456}, compared with 
results obtained from normalization point scheme}
\vskip 2mm
\begin{center}
\begin{tabular}{c|c|c|c|c|c}

N & Nickel index & & original value & corrected value & current work\\
\hline
\multirow{4}{*}{9 } & \multirow{4}{*}{e112-23-34-44-e-} & $1/\epsilon\;$ & 0.4 & -0.93333& -0.93335(9)\\
 & & $1/\epsilon^{2}$ & 1.466666 & 1.466666 & 1.466669(9) \\
 & & $1/\epsilon^{3}$& -1.066666& -1.066666 & -1.0666665(8)\\
 & & $1/\epsilon^{4}$& 0.533333& 0.533333 & 0.533333\\
\hline
\multirow{4}{*}{10 } & \multirow{4}{*}{e112-34-334-4-e-} & $1/\epsilon\;$ & 0.3563478 & 0.0316508& 0.03164(7)\\
 & & $1/\epsilon^{2}$ & 0.641666 & 0.641666 & 0.641670(8) \\
 & & $1/\epsilon^{3}$& -0.7666666& -0.7666666 & -0.7666665(5)\\
 & & $1/\epsilon^{4}$& 0.4& 0.4 & 0.4\\
\hline
\end{tabular}
\end{center}
\end{table}

Nickel index mentioned in the tables \ref{tab:G2},\ref{tab:G4}  is the most efficient way to identify diagrams (to be more precise it is adjacency list ordered in special way), one can find its definition in \cite{nickel}. The numbering of diagrams (first column) in table \ref{tab:G2} is based on papers \cite{phi412}, in table \ref{tab:G4} -- on papers \cite{phi434}.

\begin{table}[h!]
\caption{\label{tab:G4} Diagrams from four-point Green function calculated in \cite{phi434} and corrected in \cite{phi456}, compared with results obtained from normalization point scheme (``s.c.'' = symmetry coefficient)}
\vskip 2mm
\begin{center}
\begin{tabular}{c|c|c|c|c|c}
N & Nickel index & & original value & corrected value & current work\\
\hline
\multirow{2}{*}{2 } & \multirow{2}{*}{e123-e23-45-45-e5-e-} & s.c. &  3 &  3/2&  3/2\\
 & & $1/\epsilon$ & 20.807147 & 20.807147 & 20.807141(50) \\ 
\hline
\multirow{2}{*}{24} & \multirow{2}{*}{e112-34-e35-45-e5-e-} & $1/\epsilon\;$ & 14.246950 & 5.860538& 5.860525(14)\\
 & & $1/\epsilon^{2}$ & -4.14771102 & -4.14771102 & -4.1477109(10) \\
\hline
\multirow{3}{*}{32} & \multirow{3}{*}{ee12-234-34-45-5-ee-} & $1/\epsilon\;$ & 12.36505 & 1.995772& 1.99578(20)\\
 & & $1/\epsilon^{2}$ & -16.734897 &  -16.734897 &  -16.734892(6) \\
& & $1/\epsilon^{3}$ & 11.539746 & 11.539746 & 11.539747(1) \\
\hline
\multirow{4}{*}{32} & \multirow{4}{*}{ee12-ee3-445-455-5--} & $1/\epsilon\;$ & -1.70502 & -1.12169& -1.12165(10)\\
 & & $1/\epsilon^{2}$ & 2.133333 &  2.133333 &  2.133338(20) \\
& & $1/\epsilon^{3}$ & -1.333333 & -1.333333 & -1.333333(1) \\
& & $1/\epsilon^{4}$ & 0.533333 & 0.533333 & 0.533333 \\
\hline
\end{tabular}
\end{center}
\end{table}

It can be stated that discussed approach shows its efficiency and can be used in wide range of models.

\section*{Acknowledgments}
The authors are grateful to N.V.~Antonov, K.G.~Chetyrkin, M.~Hnatic, A.L.~Kataev and  D.I.~Kazakov for helpful discussions and advices, to ACAT'13 Organizing Committee, and in particular Jian-Xiong Wang (IHEP, Beijing) for support and hospitality, to the members of ACAT'13 Int. Advisory Committee, and in particular  A.B.~Arbuzov (JINR, Dubna) and A.L.~Kataev (INR RAS) for the support to present this talk on ACAT'13.
The work was supported in part by the Russian Foundation for Fundamental Research (project 12-02-00874-a).
Research was carried out using computer resources provided by Resource Center "Computer Center of SPbU" (http://cc.spbu.ru)

\end{document}